# Proximity Enhanced Quantum Spin Hall State in Graphene

Liangzhi Kou[1*], Feiming Hu[1*], Binghai Yan,[2,3] Tim Wehling,[1] Claudia Felser,[2] Thomas Frauenheim,[1] and Changfeng Chen[4]

[1]Bremen Center for Computational Materials Science, University of Bremen, Am Falturm 1, 28359, Bremen, Germany

[2]Max Planck Institute for Chemical Physics of Solids, NoethnitzerStr. 40, 01187 Dresden, Germany

[3]Max Planck Institute for Physics of Complex Systems, NoethnitzerStr. 38, 01187 Dresden, Germany

[4]Department of Physics and Astronomy and High Pressure Science and Engineering Center, University of Nevada, Las Vegas, Nevada 89154, United States

**Abstract**

Graphene is the first model system of two-dimensional topological insulator (TI), also known as quantum spin Hall (QSH) insulator. The QSH effect in graphene, however, has eluded direct experimental detection because of its extremely small energy gap due to the weak spin-orbit coupling. Here we predict by ab initio calculations a giant (three orders of magnitude) proximity induced enhancement of the TI energy gap in the graphene layer that is sandwiched between thin slabs of $Sb_2Te_3$ (or $MoTe_2$). This gap (1.5 meV) is accessible by existing experimental techniques, and it can be further enhanced by tuning the interlayer distance via compression. We reveal by a tight-binding study that the QSH state in graphene is driven by the Kane-Mele interaction in competition with Kekulé deformation and symmetry breaking. The present work identifies a new family of graphene-based TIs with an observable and controllable bulk energy gap in the graphene layer, thus opening a new avenue for direct verification and exploration of the long-sought QSH effect in graphene.

Corresponding author.

E-mail address: kouliangzhi@gmail.com (L. Kou) and feiminghu@gmail.com (F. Hu).

## 1. Introduction

The discovery of topological insulators (TIs) represents one of the most important advances in physics and materials science in recent years [1, 2]. Of particular interest are the two-dimensional (2D) TIs that exhibit quantum spin Hall (QSH) effect with gapless spin-polarized edge states that can sustain dissipationless current [3–5]. Kane and Mele first predicted QSH effect in graphene [3], which has since become a prototypical model system for 2D TIs [1, 2]. This intriguing effect, however, has eluded direct experimental detection because the bulk energy gap of graphene is extremely small ($10^{-3}$ meV) due to its weak spin-orbit coupling (SOC) [6, 7]. Subsequent studies have proposed to enhance the SOC in graphene by heavy adatom doping [8, 9] or to explore other materials [1, 2]. Recent work reported synthesis of high-quality ultrathin $Bi_2Se_3$ grown on a pristine graphene substrate using a vapor-phase deposition method [10, 11]. It was argued that the proximity effect would significantly enhance the SOC in graphene [12, 13]. Such heterostructures, however, are topologically trivial insulators despite the enhanced SOC, thus are not expected to exhibit QSH effect. Turning graphene into a true 2D TI with a measurable bulk energy gap remains an outstanding challenge.

In this paper, we propose a new strategy to solve this challenging problem by constructing quantum well (QW) structures where a graphene layer is sandwiched between properly selected materials with strong SOC. First-principles calculations show that graphene sandwiched between two single quintuple-layers (QLs) of $Sb_2Te_3$ exhibits a proximity enhanced band gap of 1.5 meV at the graphene Dirac point, which is a three-orders-of-magnitude enhancement over its intrinsic gap. The obtained QSH state is characterized by the topologically nontrivial $Z_2$ index and an explicit demonstration of the presence of topological edge states. Moreover, compressing the interlayer distance can effectively tune and further enhance the gap. QSH states are also obtained by sandwiching graphene between thin slabs of $MoTe_2$ where material and structural variations do not change the fundamental physics. We reveal by a tight-binding study that the topological phase transition is dominated by the enhanced Kane-Mele interaction in competition with the Kekulé deformation and symmetry breaking by interlayer stacking patterns. The present results suggest an effective approach to the study of QSH effect in graphene via proximity enhanced SOC in a new family of 2D TI structures.

## 2. Methods and Model

First-principles calculations based on the density functional theory (DFT) were carried out using the Vienna Ab Initio Simulation Package (VASP) [14]. The exchange correlation interaction was treated within the local-density approximations (LDA) [15]. Projector augmented wave potentials were employed to represent the ions. The Brillouin zone integration was sampled by a $10\times10\times1$ k-grid mesh. An energy cutoff of 300 eV was chosen for the plane wave basis. The SOC was included in the calculations. To describe the van der Waals (vdW) type interaction between graphene and the TI surface, we employed a semiempirical correction by Grimme's method [16]. We show in Fig. 1 the proposed QW structure where a graphene layer is sandwiched between two $Sb_2Te_3$ QLs. This structural construction preserves the inversion symmetry, and there is a prefect lattice match between graphene ($\sqrt{3}\times\sqrt{3}$ supercell) and $Sb_2Te_3$ (primitive unitcell). We have chosen the experimental in-plane lattice constant of 4.26 Å for $Sb_2Te_3$ and 2.46 Å for graphene. To obtain the most stable configuration and the stacking pattern between graphene and $Sb_2Te_3$, an extensive energetic study has been performed. We began with the initial stacking pattern with the Te atoms at the hollow center of hexagon carbon rings (Fig. 1a), and monitored the total energy variation when moving cladding layer $Sb_2Te_3$ relative to the graphene layer. The obtained total energies for the hollow stacking patterns are lowest among all the possible configurations (see Fig. 1c), which is about 30 meV/unit-cell lower than the stacking pattern with Te atom on top of carbon. Below we will focus our study on the most stable hollow stacking pattern.

## 3. Results and Discussions

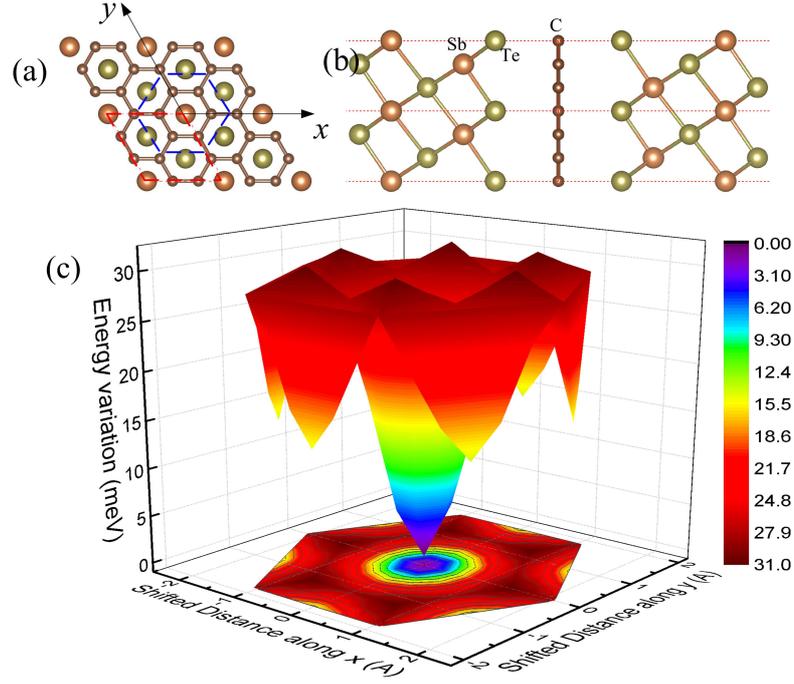

FIG. 1: (color online) The proposed construction of the $Sb_2Te_3$/Graphene/$Sb_2Te_3$ quantum well structure. Panel(a) and (b) show the top and side view, respectively, of the stacking pattern with Te atoms at the hollow center of the hexagon carbon rings, which is the most stable configuration. The red dashed lines in (a) indicate the unit cell. Panel (c) shows the energy variation when one of the cladding layers is shifted away from the equilibrium position with the shifted area indicated by the blue dashed lines in (a).

For a freestanding graphene layer with a $\sqrt{3}\times\sqrt{3}$ supercell, the two Dirac valleys at the K and K' points are folded onto the $\Gamma$ point in the Brilloun zone (Fig. 1b), and the Dirac bands are four-fold degenerate (including the spin degeneracy). While bulk $Sb_2Te_3$ is a well-defined 3D TI with a single Dirac cone, the interaction between the states on opposite surfaces of the single QL $Sb_2Te_3$ opens a gap that removes the helical Dirac cone, converting the QL to a normal 2D insulator [17–19]. The electronic band structure of the QW (Fig. 2) contains a pair of Dirac cones near the Fermi level, which are exclusively contributed by graphene according to a band state analysis (red dotted states). Without considering the SOC, the graphene Dirac cones remain four-fold degenerate as in the freestanding case, but a sizable band gap ($E_g$ = 22 meV) has opened up (see Fig. 2a). Although the QW structure preserves the AB sublattice symmetry in graphene, there is a symmetry breaking by the $Sb_2Te_3$ slabs. From Fig. 1a, one can see that the three nearest neighbors of a carbon site are no longer equivalent, and among the three C-C bonds connecting these nearest neighbors, two of them each crosses a Sb-Te bond in the adjacent $Sb_2Te_3$ layer and one crosses a Te-Te bond. This type of symmetry breaking is described by the Kekulé deformation, which leads to the opening of a band gap [20]. The strong SOC in Sb2Te3 is expected to produce considerable proximity-induced SOC in the graphene layer, thus influencing the electronic properties.

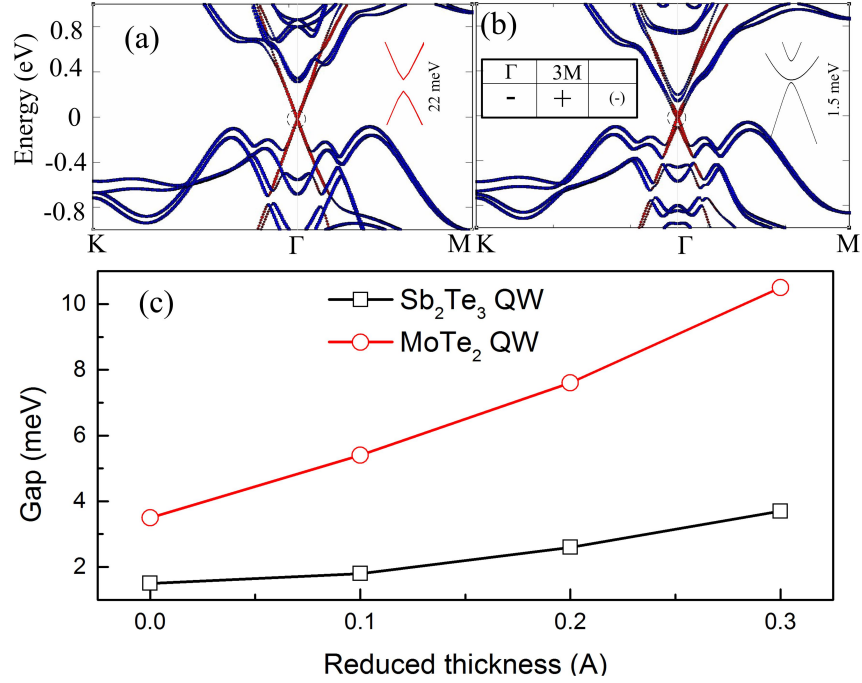

FIG. 2: (color online) Calculated electronic band structure of the $Sb_2Te_3$/Graphene/$Sb_2Te_3$ QW (a) without and (b) with the SOC at the equilibrium position; the inset in (b) shows a zoomed-in view of the bands near the Fermi level. The blue and red dots are the states from $Sb_2Te_3$ and graphene respectively. Presented in (c) is the band gap variation as a function of the interlayer distance.

Our calculations show that, when the SOC is switched on, the Dirac bands of the graphene layer remain inside the gap of the $Sb_2Te_3$ slabs, but there is a band splitting near the Fermi level, which is attributed to the fact that the induced SOC by adjacent $Sb_2Te_3$ slabs has an effective repulsive interaction for the folded Dirac bands at the Γ point. Another profound effect of the SOC is a significant shift of the graphene Dirac bands toward the Fermi level (see Fig. 2b). It brings the two Dirac cones into close proximity and, as we will show below, this produces a topological phase transition in the graphene layer, rendering it a true 2D TI with a band gap of 1.5 meV, which represents a three-orders-of-magnitude enhancement over the intrinsic gap of graphene. While the $Sb_2Te_3$ QLs play a key role in generating the proximity enhanced TI band gap in the graphene layer, their electronic bands do not mix near the Fermi level, making the QW structure a desirable (i.e., "clean") system for study of the QSH effect in graphene.

The fundamental TI nature is captured by the parity of its electronic states measured by the $Z_2$ index [21]. We have calculated the $Z_2$ index for the $Sb_2Te_3$/Graphene/$Sb_2Te_3$ QW structure by directly evaluating the parity eigenvalues following the Fu-Kane criteria [21], and found that the product of all the valence band parities at the two time-reversal invariant k-points, $\Gamma$ and M, are "-", indicating $Z_2$=1, thus confirming the QW structure is a 2D TI. One can evaluate the 2D TI state from a simple band inversion picture. Without the SOC, the two highest valence bands and two lowest conduction bands exhibit "+" and "-" parities at the $\Gamma$ point, respectively. With the SOC included, the highest valence band and the lowest conduction band swap their parities, and this inversion of bands with opposite parities induces the nontrivial $Z_2$ phase. Since the enhanced SOC in graphene is induced by the proximity effect, that is, the hybridization between the graphene and Sb2Te3 layers, it is expected that a reduction in the interlayer distance would lead to further enhanced SOC. In experiments, this can be achieved by applying a compressive pressure perpendicular to the QW layers. Results in Fig. 2(c) show that reducing the interlayer distance below the equilibrium value by external compression leads to additional increase of the band gap $E_g$ that can reach 3.5 meV at $\Delta d$ = -0.3Å. This suggests an effective method to controllably tune and increase the TI band gap of graphene.

To unveil the mechanism and the underlying physics, we performed a tight-binding (TB) study on graphene and adjusted the TB parameters to simulate the effects of the bonding environment in the QW structure and the external compression. We write the Kane-Mele Hamiltonian [3, 22] for the graphene lattice as,

$$H = -\sum_{<ij>} t_{ij} c_i^* c_j + \sum_{\ll ij \gg} i\lambda v_{ij} s^z c_i^* c_j$$

The first term in Eq. (1) is the hopping matrix, and in a nearest-neighbor model for freestanding graphene t = 2.60 eV [23]. In the presence of the $Sb_2Te_3$ films, however, the $t_{ij}$ in the graphene layer becomes site-dependent due to a symmetry breaking caused by the Kekulé deformation. The second term is the SOC. In pristine graphene λ is about $10^{-3}$ meV, but here it is significantly enhanced by the proximity effect; $s^z$ is the z-compound of the Pauli matrix. The parameters $v_{ij}$ =$-v_{ji}$=±1 depend on the direction of electron hopping from site i to j: $v_{ij}$=+1 if the electron makes a left turn to a nearest-neighbor bond, and it is -1 if it makes a right turn [3].

The Hamiltonian in the momentum space for one of the spin species can be expressed as

$$H_{k\uparrow} = \begin{pmatrix} 0 & t & i\lambda\beta_1 & t'e^{i(k_y-k_x)} & -i\lambda\beta_1 & t \\ t & 0 & t & i\lambda\beta_2 & t'e^{ik_y} & i\lambda\beta_3 \\ -i\lambda\beta_1^* & t & 0 & t & i\lambda\beta_3 & t'e^{ik_x} \\ t'e^{-i(k_y-k_x)} & -i\lambda\beta_2^* & t & 0 & t & i\lambda\beta_1^* \\ i\lambda\beta_1^* & t'e^{-ik_y} & -i\lambda\beta_3^* & t & 0 & t \\ t & i\lambda\beta_3^* & t'e^{-ik_x} & -i\lambda\beta_1 & t & 0 \end{pmatrix}$$

In every hexagonal unit cell, there are six sites, so after Fourier transformation performed on the original model in Eq.(1), the Hamiltonian in momentum space is a 6×6 matrix and the basis is ($c_{1k\sigma}^\dagger$, $c_{2k\sigma}^\dagger$, $c_{3k\sigma}^\dagger$, $c_{4k\sigma}^\dagger$, $c_{5k\sigma}^\dagger$, $c_{6k\sigma}^\dagger$) here σ is the index of spin. $\beta_1$= 1 +$e^{-ik_x}$ + $e^{i(k_y-k_x)}$, $\beta_2$= 1 +$e^{ik_y}$ + $e^{i(k_y-k_x)}$, $\beta_3$= 1 +$e^{ik_y}$ + $e^{ik_x}$, and it is noted that $H_{k\uparrow} = H_{k\downarrow}^*$. The difference between t' and t reflects the effect of the Kekulé deformation that describes the different bonding and hopping environment as discussed above (see Fig. 1c).

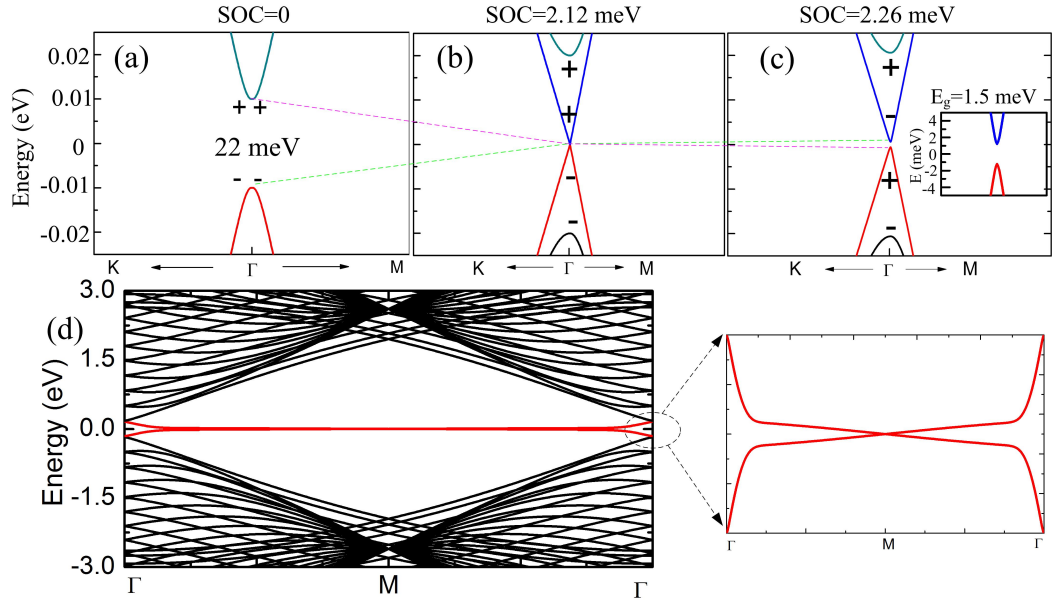

FIG. 3: (color online) Evolution of the TB band structure of the graphene lattice as the SOC parameter is tuned. (a) The band gap opening due to the Kekulé deformation without considering the SOC; the Dirac cones are degenerate, and an eigenpair analysis indicates the parities ++ and -- for the conduction and valence bands, respectively. (b) Under increasing SOC both the conduction and valence bands split, and at a critical value of SOC (2.12 meV) the gap induced by the Kekulé deformation is completely closed. (c) Further increase of the SOC reopens the gap, leading to an inverted band gap of 1.5 meV at the SOC value of 2.26 meV corresponding to the equilibrium QW structure, and the + and - signs indicate the parities of the inverted band states. (d) The tight-binding band structure of a zigzag graphene ribbon obtained using the same parameters that produced the results in (c); the helical topological edge states are explicitly demonstrated.

We first examine the effect of the Kekulé deformation on the electronic band structure without considering the SOC, and the results are shown in Fig. 3a. The asymmetric hopping parameters produce a band gap $2\Delta t$ ($\Delta t = t'-t$), and a fitting to the ab initio band gap of 22 meV gives t' = 2.611 eV; the parity eigenvalues are (++,--) for the two conduction bands and two valence bands, respectively, corresponding to a topologically trivial insulator state. Once the SOC is turned on and its strength gradually increased, the degenerate Dirac cones at the Γ point split by an effective SOC induced repulsive interaction, reducing the gap for one pair of the Dirac cones and increasing the gap for the other. When the SOC reaches 2.12 meV, the declining gap is completely closed as shown in Fig. 3b; further increasing SOC reopens the gap, leading to an inverted band gap with a new parity sequence (+-, +-) that characterizes a true 2D TI phase. At the SOC value of 2.26 meV, the inverted band gap reaches the value of 1.5 meV (Fig. 3c) predicted by the ab initio calculations in the equilibrium QW structure. These TB results offer important insights into the competing roles of the Kekulé deformation and the proximity induced SOC effect. At the graphene Dirac point, the Kekulé term opens a topologically trivial insulator band gap ($2\Delta t$), while the Kane-Mele term (SOC) produces a topologically nontrivial band gap ($6\sqrt{3}\lambda$). These two terms combine to determine the final band gap $E_g = 2\Delta t - 6\sqrt{3}\lambda$. When the SOC term is dominant and $E_g$ is negative as in the present case, it indicates an inverted band gap and a TI phase; otherwise it signifies a trivial insulator phase. Another characteristic feature of a 2D TI is the existence of 1D helical edge states. For an explicit demonstration, we have constructed a zigzag graphene ribbon that comprises 60 carbon chains with a width of 14.76 nm. We calculated the TB band structure using the parameters t = 2.60 eV, t'=2.611 eV, and λ=2.26 meV that are mapped out from fitting the ab initio band structure. We indeed obtained a pair of gapless edge states inside the 2D bulk gap as shown in Fig. 3(e). These states are localized at both edges of the ribbon, and at a given edge two counter propagating edge states exhibit opposite spin polarizations, showing their 1D helical feature (i.e., spin momentum locking).

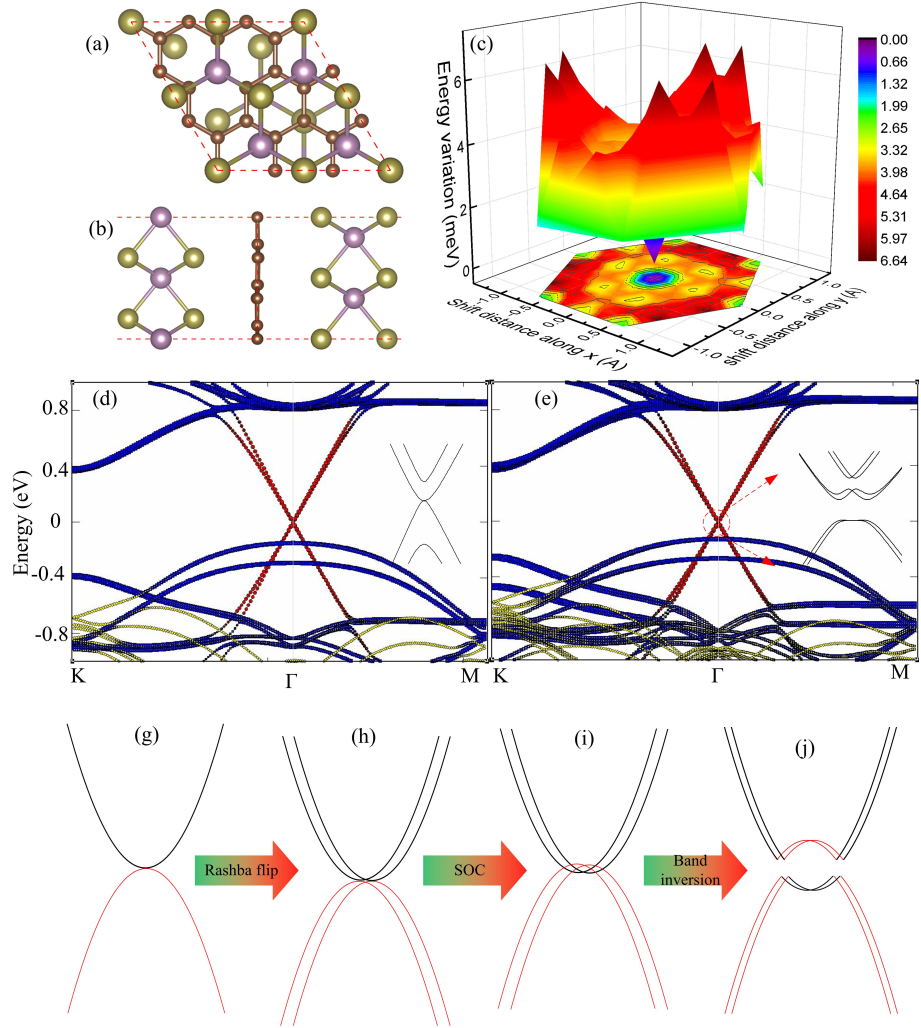

FIG. 4: (color online) (a) Top and (b) side view of the MoTe$_2$/Graphene/MoTe$_2$ QW structure, where the interlayer stacking is incommensurate. The calculated total energies as a function of relative positions are shown in (c), and the results show that the stacking pattern with the graphene layer sandwiched between AB stacked bilayer MoTe$_2$ is the most stable. The obtained electronic band structures without and with SOC are shown in (d) and (e) respectively, and the zoomed-in views of the Dirac cones are shown in the insets. The band inversion process is illustrated in panels (g) to (j).

The 2D TI state in graphene-based QW structures is not limited to the Sb$_2$Te$_3$ case. It can be generalized to other materials that share a common thread of having a band structure that can accommodate the graphene Dirac cones inside its gap and possessing a SOC that can pull these Dirac cones into close proximity to induce a topological band inversion. $Sb_2Te_3$ is 3D TI although it becomes a topologically trivial insulator in single QLs [1, 2]. Selection of a 3D TI, however, is not a prerequisite for constructing the 2D TI QW structures. Here we present a case study of graphene sandwiched between two single triple-layers (TLs) of $MoTe_2$, which is not a TI but possesses strong SOC [24]. The significantly smaller lattice constant of $MoTe_2$ (3.56 Å) precludes an interlayer stacking that preserves the AB sublattice symmetry in the graphene layer. In our structural model, we match a 3×3 graphene supercell with a 2×2 $MoTe_2$ supercell (see Fig. 4a, 4b), and the resulting structure has a relatively small lattice mismatch of 3.6% between the $MoTe_2$ and graphene layer. Similar to the $Sb_2Te_3$ QW, extensive energetic calculations were also performed to determine the most stable stacking pattern. Since AB stacked bilayer transition metal dichalcogenides are the most stable configuration [25], we used AB stacked $MoTe_2$ with graphene sandwiched inside as the initial configuration. The energetic calculations indicate that this initial configuration and stacking pattern has the lowest total energy although the energy difference between different stacking patterns are quite small, only 5 meV per supercell, see Fig. 4(c). Our ab initio band structure calculations confirm that the graphene Dirac cones, which are located inside the band gap of $MoTe_2$, remain touching each other near the Fermi level (Fig. 4d). Due to a lack of the inversion symmetry, the parity criteria for topological insulators do not apply here. However, a band gap of 3.5 meV opens when the SOC is turned on (Fig. 4e). Different from the inversion symmetry induced high band degeneracy in the $Sb_2Te_3$ QWs, here we observe an obvious band flip in the band structures where the valence and conduction bands are shifted relative to each other. This phenomenon originates from the Rashba effect driven by the asymmetric AB stacked $MoTe_2$. This SOC and Rashba effect driven band inversion process is illustrated in Fig. 4(g) to 4(j). The two bands near the Fermi level split and flip along the K-space when the Rashba effect is present. But when the proximity SOC is considered, the energies of these band states are shifted, leading to the band inversions. Such band inversions render the QW structure into a topological insulator phase. Compression pushes the $MoTe_2$ and graphene layers closer, and the resulting enhanced proximity effect leads to a significant increase of the nontrivial gap as shown in Fig. 2(c), which suggests that the gap can reach up to 10 meV when the interlayer distance is reduced by 0.3 Å. The enhanced proximity SOC in graphene has been demonstrated by recent experiments, which show that graphene acquires a spin-orbit coupling up to 17 meV when placing it on few-layer semiconducting tungsten disulphide [26]. The present results demonstrate that the 2D TI phase in QW structures is versatile in material choice and robust against variations in the interlayer stacking patterns. The insensitivity to material and structural details for the topological phase transition bodes well for experimental synthesis and exploration of QSH state in graphene.

## 4. Conclusion

In summary, we have identified by ab initio calculations a new family of QW structures that exhibit 2D TI state in the graphene layer that is sandwiched between thin slabs of strong-SOC materials. The topological phase transition in the graphene layer is driven by proximity enhanced SOC effect that competes with the influence of the broken translational symmetry caused by anisotropic or incommensurate interlayer stacking. The resulting TI band gap is three orders of magnitude larger than that in pristine graphene, making it accessible to experimental probes. This gap can be tuned and further enhanced by compressing the interlayer distance between graphene and strong-SOC materials. These QW structures, which can be synthesized by available epitaxial techniques, offer a promising new template for direct verification and systematic study of the quantum spin Hall effect in graphene.

## Acknowledgement

Computation was carried out at HLRN Berlin/Hannover (Germany). L.K. acknowledges financial support by the Alexander von Humboldt Foundation of Germany. C.C. was supported by the U.S. Department of Energy under Cooperative Agreement DE-NA0001982. B.Y. and C.F. acknowledge financial support from the European Research Council Advanced Grant (ERC 291472). F.H. acknowledges a Postdoctoral Research Fellowship from University of Bremen.

# References

[1] Hasan MZ, Kane CL, Colloquium: Topological Insulators, Rev. Mod. Phys. 2010; 82:3045-3067.

[2] Qi XL, Zhang SC, Topological Insulators and Superconductors, Rev. Mod. Phys. 2011; 83:1057-1110.

[3] Kane CL, Mele EJ, Quantum Spin Hall Effect in Graphene, Phys. Rev. Lett. 2005; 95:226801-4.

[4] Bernevig BA, Hughes TL, Zhang SC, Quantum Spin Hall Effect and Topological Phase Transition in HgTe Quantum Wells, Science 2006; 314:1757-1761.

[5] Roth A, Brüne C, Buhmann H, Molenkamp LW, Maciejko J, Qi XL, Zhang SC, Nonlocal Transport in the Quantum Spin Hall State, Science 2009; 325:294-297.

[6] Yao Y, Ye F, Qi XL, Zhang SC, Fang Z, Spin-orbit Gap of Graphene: First-principles Calculations, Phys. Rev. B 2007; 75:041401(R)-4.

[7] Min H, Hill JE, Sinitsyn NA, Sahu BR, Kleinman L, MacDonald AH, Intrinsic and Rashba Spin-orbit Interactions in Graphene Sheets, Phys. Rev. B 2006; 74:165310-5.

[8] Weeks C, Hu J, Alicea J, Franz M, Wu R, Engineering a Robust Quantum Spin Hall State in Graphene via Adatom Deposition, Phys. Rev. X 2011; 1:021001-15.

[9] Castro Neto AH, Guinea F, Impurity-Induced Spin-Orbit Coupling in Graphene, Phys. Rev. Lett. 2009; 103: 026804-4.

[10] Dang W, Peng H, Li H, Wang P, Liu Z, Epitaxial Heterostructures of Ultrathin Topological Insulator Nanoplate and Graphene, Nano Lett. 2010; 10:2870-2876.

[11] Song CL, et al, Topological Insulator $Bi_2Se_3$ Thin Films Grown on Double-layer Graphene by Molecular Beam Epitaxy, Appl. Phys. Lett. 2010; 97:143118-3.

[12] Jin KH, Jhi SH, Proximity-induced Giant Spin-orbit Interaction in Epitaxial Graphene on a Topological Insulator, Phys. Rev. B 2013; 87:075442-6.

[13] Liu W, Peng X, Wei X, Yang H, Stocks GM, Zhong J, Surface and Substrate Induced Effects on Thin Films of The Topological Insulators $Bi_2Se_3$ and $Bi_2Te_3$, Phys. Rev. B 2013; 87:205315-6.

[14] Kresse G, Furthmüller J, Effcient Iterative Schemes for Ab Initio Total-energy Calculations using a Plane-wave Basis Set, Phys. Rev. B 1996; 54:11169-11186.

[15] Perdew JP, Zunger A, Self-interaction Correction to Density-functional Approximations for Many-electron Systems, Phys. Rev. B, 1981; 23:5048-5079.

[16] Grimme S, Semiempirical GGA-type Density Functional Constructed with a Long-range Dispersion Correction, J. Comput. Chem. 2006; 27:1787-1799.

[17] Liu CX, Zhang HJ, Yan B, Qi XL, Frauenheim T, Dai X, Fang Z, Zhang SC, Oscillatory Crossover from Two-dimensional to Three-dimensional Topological Insulators, Phy. Rev. B 2010; 81:041307(R)-4.

[18] Lu HZ, Shan WY, Yao W, Niu Q, Shen SQ, Massive Dirac Fermions and Spin Physics in an Ultrathin Film of Topological Insulator, Phys. Rev. B 2010; 81:115407-7.

[19] Li YY, et. al., Intrinsic Topological Insulator $Bi_2Te_3$ Thin Films on Si and Their Thickness Limit, Adv. Mater. 2010; 22:4002-4007.

[20] Cyvin SJ, Kekulé Structures of Polyphenes, Monatshefte für Chemie 1982; 113:1127-1131.

[21] Fu L, Kane CL, Topological Insulators with Inversion Symmetry, Phys. Rev. B 2007; 76:045302-17.

[22] Kane CL, Mele EJ, $Z_2$ Topological Order and the Quantum Spin Hall Effect, Phys. Rev. Lett. 2005; 95:146802-4.

[23] Jung J, Zhang F, Qiao Z, MacDonald AH, Valley-Hall Kink and Edge States in Multilayer Graphene, Phys. Rev. B 2011; 84:075418-5.

[24] Zhu ZY, Cheng YC, Schwingenschlögl U, Giant Spin Orbit-induced Spin Splitting in Two-dimensional Transition-metal Dichalcogenide Semiconductors, Phys. Rev. B. 2011; 84:153402-5.

[25] He J, Hummer K, Franchini C, Stacking Effects on the Electronic and Optical Properties of Bilayer Transition Metal Dichalcogenides $MoS_2$, $MoSe_2$, $WS_2$, and $WSe_2$, Phys. Rev. B 2014; 89:075409-11.

[26] Avsar A, et. al. Spin Orbit Proximity Effect in Graphene, Nature Comm., 2014 doi:10.1038/ncomms5875.